\documentclass[aps,prl,amsmath,amssymb,nofootinbib,twocolumn, reprint]{revtex4-1}

\usepackage{graphicx}
\usepackage{dcolumn}
\usepackage{footmisc}
\usepackage[normalem]{ulem}
\usepackage{bm}
\usepackage{hyperref}
\usepackage{xcolor}
\usepackage{float}
\usepackage{hyperref}

\begin{document}
\setlength{\abovedisplayskip}{5pt}
\setlength{\belowdisplayskip}{5pt}
\setlength{\abovedisplayshortskip}{5pt}
\setlength{\belowdisplayshortskip}{5pt}
\hyphenpenalty=1050

\preprint{}

\title{New Physics versus Quenching Factors in Coherent Neutrino Scattering}

\author{Yulun Li, Gonzalo Herrera, Patrick Huber}
\affiliation{Center for Neutrino Physics, Department of Physics, Virginia Tech, Blacksburg, VA 24061, USA}

\begin{abstract}
Recent results on the Coherent Elastic Neutrino-Nucleus Scattering (CE$\nu$NS) on germanium present significant discrepancies among  experiments. We perform a combined analysis of the Dresden-II, CONUS+ and COHERENT data, quantifying the impact of quenching factor uncertainties on their CE$\nu$NS cross section measurement. No choice of quenching factor can bring these three data sets into mutual agreement, whereas the combination of COHERENT with either Dresden-II or CONUS+ agrees well albeit for very different quenching factors. We further study the quenching factor dependence on the sensitivity of these experiments to a large neutrino magnetic moment, finding that the constraints can vary by up to an order of magnitude. Our work  highlights the importance of reducing this uncertainty on quenching factors in order to probe new physics from neutrinos at the low-energy frontier.
\end{abstract}

\maketitle

\section{Introduction}

Coherent Elastic Neutrino-Nucleus Scattering (CE$\nu$NS) has been observed in recent years across
a variety of detector materials, energy ranges, and experiments. The first observation of CE$\nu$NS by COHERENT in cesium iodine (CsI) in 2017 opened a new window to test Beyond the Standard Model (BSM) interactions of neutrinos \cite{Freedman:1973yd, COHERENT:2017ipa}. The subsequent indications of coherent neutrino scattering in different materials like argon or xenon (and from different neutrino sources and energies in xenon) have allowed for a better confirmation of this effect, and its expected dependence with atomic mass \cite{COHERENT:2020iec,XENON:2024ijk}. These datasets have already been used to place constraints on new interactions of neutrinos at the detector enhanced at low-energies, where the CE$\nu$NS cross section dominates over elastic neutrino-electron scatterings. Prominent examples are given by large neutrino magnetic moments \cite{Billard:2018jnl,Miranda:2019wdy,A:2022acy,DeRomeri:2024hvc}, the neutrino charge radius or anapole moment \cite{Cadeddu:2018dux,AtzoriCorona:2022qrf,Khan:2022bel,Khan:2022jnd,Herrera:2024ysj}, and new light mediators coupled to quark and neutrinos \cite{Farzan:2018gtr,Denton:2018xmq,AristizabalSierra:2022axl,Denton:2022nol,Blanco-Mas:2024ale, DeRomeri:2024iaw, Tang:2024prl}.
\begin{figure*}[t!]
    \centering
    \includegraphics[width=0.49\linewidth]{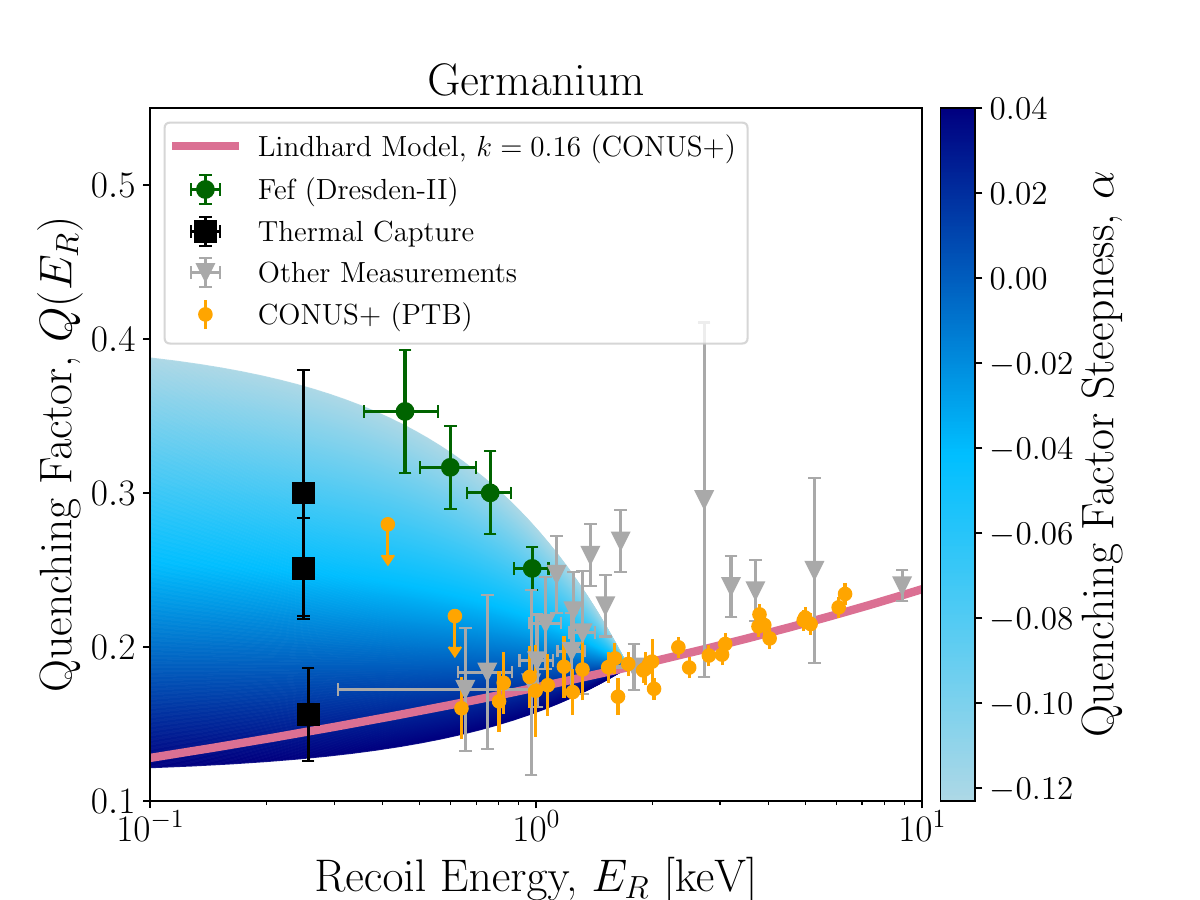}
    \includegraphics[width=0.49\linewidth]{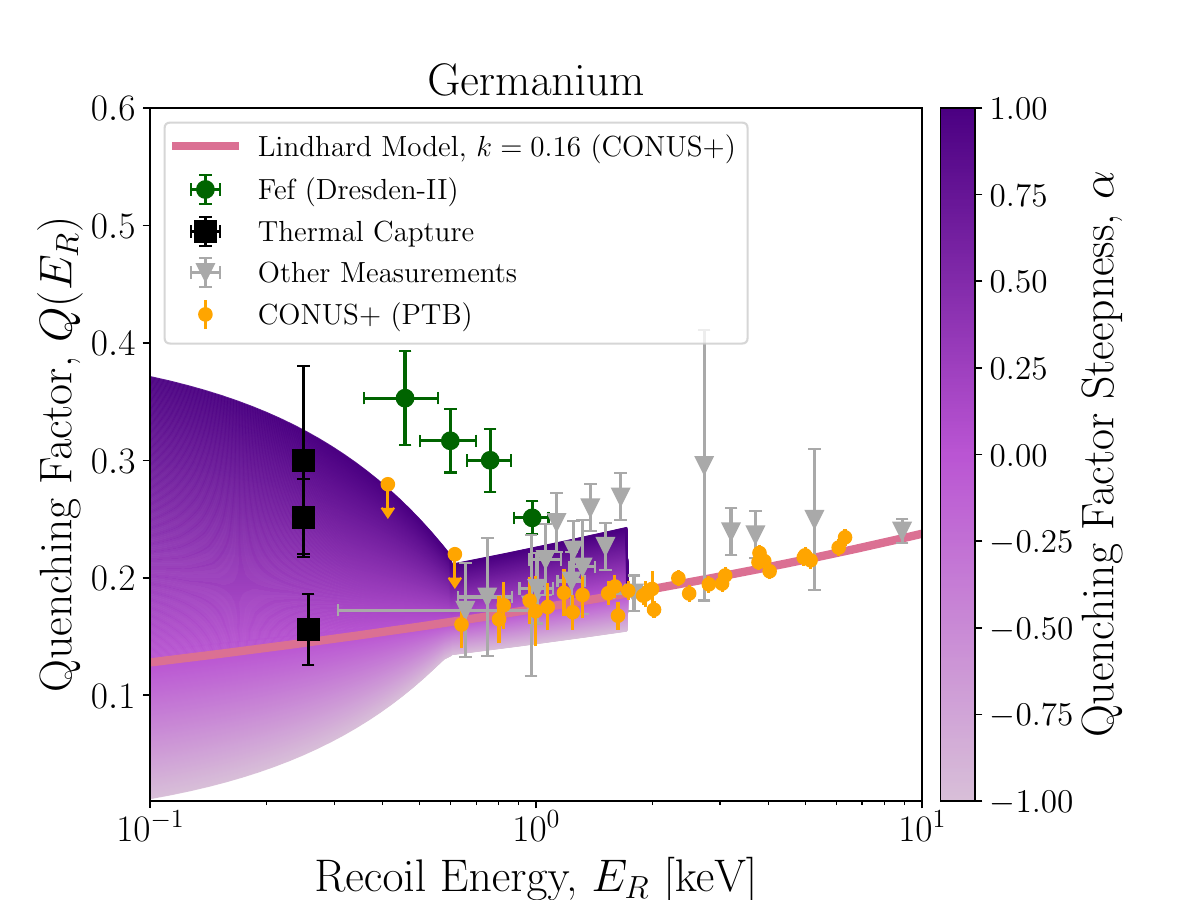}
    \caption{\textit{Left panel:} Parametrization of the germanium quenching factor at sub-keV energies from \ref{eq:qf_parametrization}, interpolating between the Lindhard model (solid pink) and Fef measurements (green data points) \cite{Collar:2021fcl}. We further show various measurements obtained via thermal neutron capture (black data points) from \cite{Collar:2021fcl} (highest quenching factor), from \cite{Kavner:2024xxd} (intermediate), and from \cite{PhysRevC.4.125} (lowest).  The Lindhard model was used at CONUS+ \cite{Ackermann:2025obx}, while the Fef measurements were used at Dresden-II \cite{Colaresi:2022obx}. We further show in orange color recent measurements and upper limits on the quenching factor from CONUS+ in collaboration with the German National Metrology Institute (PTB) \cite{Bonhomme:2022lcz}. For comparison, we also add complementary measurements of the germanium quenching factor from \cite{PhysRevC.4.125,Barbeau:2007qi, MESSOUS1995361, PhysRevA.11.1347}. Our parametrization of the quenching factor uncertainty is shown as a color gradient interpolating between two extremal steepness values at recoil energies below $E_R \leq 1.73$ keV, confer Eq. \ref{eq:qf_parametrization}. \textit{Right panel:} Parametrization of the germanium quenching factor at sub-keV energies from Eq.\ref{eq:qf_parametrization_2}, interpolating between the upper limits from CONUS+ (PTB) and the lower systematic uncertainty in their measurements \cite{Bonhomme:2022lcz}. This parametrization is more conservative than the one from Eq. \ref{eq:qf_parametrization} and does not cover the Fef measurements, however, it does cover the large uncertainties at the lowest energies measured with thermal neutron capture.}
    \label{fig:quenching_parametrization}
\end{figure*}

CE$\nu$NS with reactor neutrinos produces much lower energy nuclear recoils and its experimental observation therefore has proven to be a considerable challenge. The lower energy transfer makes reactor CE$\nu$NS more sensitive to massless or nearly massless messengers of new physics, which often phenomenologically are akin to a neutrino magnetic moment. Similarly, low recoil energies will be encountered by Dark Matter direct detection experiments aiming for the mass range below 10\,GeV. It is generally acknowledged that for detectors using ionization or scintillation, quenching of the recoil signal has to be carefully and accurately accounted for. A simple theoretical description of quenching in ionization detectors is given by the mean-field Lindhard model \cite{osti_4701226}. At recoil energies above $\sim$1\,keV this is considered to be an adequate description. However the recoil energies relevant for reactor neutrinos are well below that energy. In Fig.~\ref{fig:quenching_parametrization} we summarize the data obtained on the quenching factor in germanium and confront it with the Lindhard prediction. The spread of the data is remarkable and increases towards lower energies. Measuring low-energy quenching factors is very challenging and there are two main methods: The first method consists in using a beam of neutrons of known energy and measuring the scattering angle on an event-by-event basis. This allows to use kinematics to determine the recoil energy and hence, we call this the kinematic method. The difficulty here is to ensure that the impinging neutron beam has a sufficiently low energy and is truly mono-chromatic, even a small admixture of higher energy neutrons can significantly bias the results towards larger quenching factors. The second method relies on capture of thermal neutrons, where the capture leaves the daughter nucleus in an excited state with subsequent gamma emission. This gamma emission causes a well-defined recoil since the gamma energies and intensities are well known. The challenge for the thermal capture method arises from low statistics and significant gamma backgrounds.

The Dresden-II collaboration recently claimed an excess of events compatible with CE$\nu$NS \cite{Colaresi:2022obx}. Such an excess is furthermore compatible with the Standard Model expected cross section for CE$\nu$NS, if the iron filter (Fef) measurement of the quenching factor in germanium from \cite{Collar:2021fcl} is assumed \footnote{The measurement from \cite{Collar:2021fcl} relies on a thick iron (Fe) layer that is placed between the neutron source and the detector. Iron has has a number of resonances and anti-resonances in its scattering cross section. One of these anti-resonances sits at 24\,keV and thus a neutron beam that has scattered many times in iron effectively will be  nearly monoenergetic around 24\,keV. This monochromatic neutron flux then recoils at the detector, and the quenching factor is extracted. Similar filters can be built from a range of isotopes to create beams of different energy.}. On the other hand, the CONUS and TEXONO collaborations did not report such excess \cite{CONUSCollaboration:2024oks,TEXONO:2024vfk} when considering the quenching factor predicted in the Lindhard model \cite{osti_4701226}. Only if the parameter $k$ in the Lindhard model exceeds $k \gtrsim 0.25$, these experiments could become sensitive to CE$\nu$NS. The COHERENT collaboration further reported a measurement of CE$\nu$NS at a germanium detector compatible with the SM, although at larger recoil energies than reactor neutrino experiments, where the uncertainties in the quenching factor are less relevant \cite{Adamski:2024yqt}. Recently, the CONUS+ collaboration, with lower energy threshold and larger exposure than CONUS, claimed an excess of events arising from CE$\nu$NS, compatible within 2$\sigma$ with the Standard Model expectation when assuming the Lindhard model at low energies \cite{Ackermann:2025obx}. At first sight, the recent measurement from CONUS+ seems incompatible with the measurement from Dresden-II.
The excess of events at Dresden-II is attributed by part of the community to unidentified backgrounds rather than to CE$\nu$NS, since the Fef quenching factor measurements are discrepant with various other experimental measurements at low energies \cite{PhysRevC.4.125,Barbeau:2007qi,MESSOUS1995361, PhysRevA.11.1347}. However, the Iron filter and thermal capture experimental measurements from \cite{Collar:2021fcl}, and recently \cite{Kavner:2024xxd}, must be addressed seriously, and be included as a source of uncertainty in analyses of current experiments of the SM CE$\nu$NS scattering cross section and constraints in new physics scenarios. It is particularly interesting that the most recent thermal capture measurement from \cite{Kavner:2024xxd} is in better agreement with the thermal capture measurement from \cite{Collar:2021fcl} rather than with \cite{PhysRevC.4.125}, indicating that the quenching factor of germanium at sub-keV energies could be larger than commonly believed.

Several works have been dedicated to constrain BSM properties of neutrinos by making use of the recent hints and upper limits on CE$\nu$NS from reactor neutrino experiments based in germanium detectors, such as Dresden-II \cite{Majumdar:2022nby,AristizabalSierra:2022axl,Coloma:2022avw}, CONUS \cite{CONUS:2021dwh}, and CONUS+ \cite{Chattaraj:2025fvx, DeRomeri:2025csu,Alpizar-Venegas:2025wor,AtzoriCorona:2025ygn}. These studies focused on a concrete choice of the quenching factor of germanium at low-energies, either based on the Fef measurement or on the Lindhard model. However, this choice crucially impacts not only the agreement of these datasets with the CE$\nu$NS expected cross section, but also the reported upper limits on new physics. In particular, scenarios where the scatterings cross section of neutrino off nuclei is enhanced at low energies (such as a magnetic moment or new light mediators) are particularly affected by quenching factor uncertainties. Here we address this uncertainty by parametrizing the quenching factor as a power-law function that interpolates between the Lindhard model and the Fef measurement, and assess the compatibility of Dresden-II, CONUS+ and COHERENT data sets for all possible values of the quenching factor. We perform such analysis on the ratio of the CE$\nu$NS scattering cross section vs the SM expectation, and on the electron neutrino magnetic moment.

\begin{figure*}[t!]
		\centering
        \includegraphics[width=0.49\textwidth]{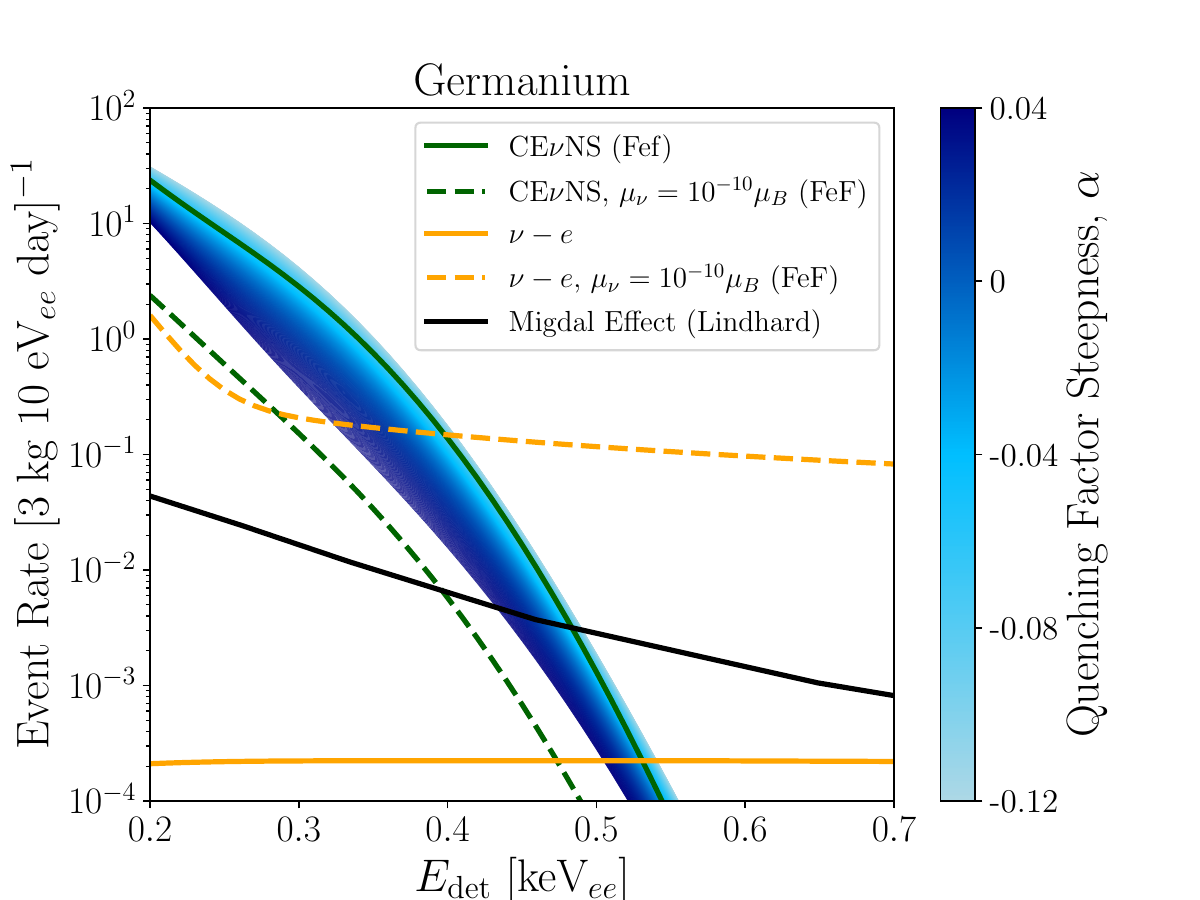}
        \includegraphics[width=0.49\textwidth]{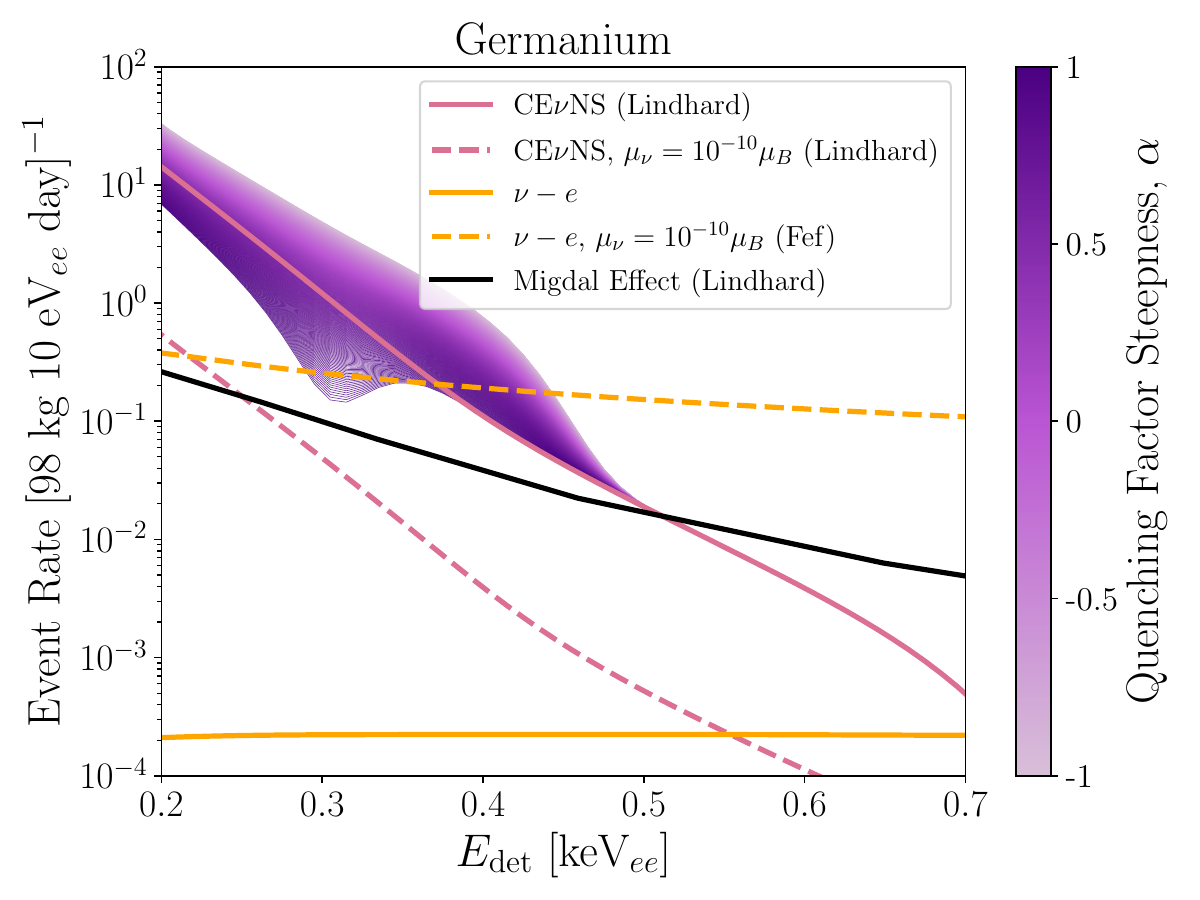}
		\caption{\textit{Left panel}: Scattering rate in terms of reconstructed ionization energy in germanium, using the resolution function and normalized to the exposure of Dresden-II. The rate is shown in the Standard Model, via electron-neutrino scattering (solid orange line), CE$\nu$NS (solid dark green line), Migdal effect (solid black line). We further show the signal induced by a large neutrino magnetic moment from neutrino-electron scattering (dashed orange) and CE$\nu$NS (dashed green line). These rates were calculated for the quenching factor measurement from Fef. The uncertainty in the scattering rate from the quenching factor modelling is indicated for CE$\nu$NS in the Standard Model as a blue band, in analogy to Figure \ref{fig:quenching_parametrization}. \textit{Right plot:} Analogous plot for the quenching factor parametrization from \ref{eq:qf_parametrization_2}, normalization fixed by CONUS+, and central benchmarks for the Lindhard model.}
        \label{fig:recoil_rate}
\end{figure*}

The paper is organized as follows: in the first section, we describe the formalism to compute the ionization induced by reactor neutrinos in germanium detectors via CE$\nu$NS, the Migdal effect and elastic neutrino-electron scattering. We describe the calculation in the SM and via a magnetic moment. Further, we describe two possible parametrizations of the quenching factor at sub-keV recoil energies, one of them accounting for values as large as those from the Fef measurement \cite{Collar:2021fcl}, and another one restricted to the (smaller) uncertainties reported in \cite{Bonhomme:2022lcz}. In the second section we derive sensitivity contours on the CE$\nu$NS cross section at Dresden-II and CONUS+ as a function of the quenching factor steepness in our two proposed parametrizations, and include the COHERENT fit at Dresden-II and CONUS+ independently. In the third section we further derive the sensitivity of Dresden-II and CONUS+ to an effective neutrino magnetic moment in terms of the quenching factor steepness. Finally, we present our conclusions.
\section{Ionization rates from CE$\nu$NS in germanium}
The cross section for CE$\nu$NS in terms of nuclear recoil energy $E_{R}$ reads \cite{Freedman:1973yd}
\begin{align}
    \frac{d\sigma}{dE_R}
        = \frac{G_F^2 m_A}{\pi} Q_V^2 \bigg(1 - \frac{m_A E_{R}}{2 E_\nu^2}\bigg)|F(E_R)|^2 \,,
    \label{eq:xsec-sm-7s}
\end{align}
where $m_A$ is the mass of the target nucleus and $Q_V$ is its vector charge, given by
\begin{align}
    Q_V = (g_{p V} Z + g_{n V} N) \,.
\end{align}
Here, $F(E_R)$ denotes the nuclear form factor, which we follow from \cite{ENGEL1991114,Coloma:2020nhf}. Alternative parametrizations of the nuclear form factor have been proposed in the literature, see \textit{e.g} \cite{Klein:1999qj}, but their impact at the energies of interest in this work are negligible. In the SM the neutral current vector couplings are $g_{p V} = \frac{1}{2} - 2\sin ^2 \theta_W$ and $g_{n V} = -\frac{1}{2}$, with $\theta_W$ the Weinberg angle. 

We will focus first on reactor antineutrino experiments. The recoil rate (in detectable energy $E_{\rm det}$ with units keV$_{ee}$) induced at a germanium detector is given by \cite{Coloma:2022avw}
\begin{equation}
\begin{aligned}
    \frac{\mathrm{d} R}{\mathrm{~d} E_{\rm det}} &= N_{\mathrm{T}} \int_{E_{\nu}^{\min}}^{\infty} \mathrm{d} E_{\nu}  \\
    &\quad \times \int_{E_{R}^{\min}}^{E_{R}^{\rm max}} \mathrm{d} E_R \frac{\mathrm{~d} \Phi_{\bar{\nu}_e}}{\mathrm{~d} E_{\nu}} \frac{\mathrm{~d} \sigma}{\mathrm{~d} E_R} \lambda(E_{\rm det},E_R),
\end{aligned}
\end{equation}
where the minimum neutrino energy necessary to induce a recoil $E_{R}$ reads (in the limit $m_A \gg E_{R}$)
\begin{align}
E_\nu^{\min }=\sqrt{m_A E_{\mathrm{nr}} / 2}.
\end{align}
Here, $N_T=2.43 \times 10^{25}$ is the number of germanium nuclei, $\mathrm{d} \Phi_{\bar{\nu}_e} / \mathrm{d} E_{\nu}$ is the reactor electron antineutrino spectrum, which we take from a combination of \cite{Vogel:1989iv} and \cite{Huber:2011wv, PhysRevC.83.054615,DayaBay:2022eyy}, and $\lambda(E_{\rm det},E_R)$ is an energy resolution function, which involves the quenching factor $Q(E_R)$ as
\begin{align}
\lambda(E_{\rm det}, E_R) = & \left(\frac{2}{1+\mathrm{Erf}\left(\frac{Q(E_R)E_R}{\sqrt{2} \sigma(E_R)}\right)}\right) 
\nonumber \\
& \times \frac{1}{\sqrt{2 \pi} \sigma(E_R)} 
e^{-\frac{\left(E_{\rm det}-Q(E_R)E_R\right)^2}{2 \sigma(E_R)^2}},
\end{align}
\begin{table*}[t!]
    \centering
    \begin{ruledtabular}
    \begin{tabular}{lcccc}
                                          & \multicolumn{1}{c}{Dresden-II} &
                                          \multicolumn{1}{c}{CONUS+} &
                                           \multicolumn{1}{c}{COHERENT}  \\  \hline \\
        Exposure [kg days]             &        281.92                                    &     282               &       4498.3                   \\
        Observed $\nu$ events     &          128 $\pm$ 11.3                                             &                        395 $\pm$ 106 &  $20.6 \pm 7.1$                         \\
        Expected $\nu$ events   &                      121 $\pm$ 11      &                347 $\pm$ 34  &                             $35.1 \pm 3.6$                      \\ \hline \\
        Energy resolution, $\sigma(E_R) \, [\rm eV]$                  &
$\sqrt{4471+0.325Q(E_R)E_R} $                       &     $\sqrt{900+0.018Q(E_R)E_R}$                                                  &          $580 \sqrt{Q(E_R)E_R}$               \\
        Energy threshold, $E_R^{\rm min} [ \rm eV]$                              & $200/Q(E_R)$                     &   $170/Q(E_R)$                          &                $1500/Q(E_R)$          \\
    \end{tabular}
    \end{ruledtabular}
    \caption{Experimental parameters used in our analysis. In order: Exposure in kg days, total number of observed neutrino-induced events in the region of interest, total number of expected neutrino-induced events in the region of interest, resolution function, and threshold recoil energy of the experiment. We show the data from Dresden-II \cite{Colaresi:2022obx}, CONUS+ \cite{Ackermann:2025obx} and COHERENT \cite{Adamski:2024yqt}.}
    \label{tab:exp-params}
\end{table*}
\begin{figure*}[t!]
    \centering
    \includegraphics[width=0.495\linewidth]{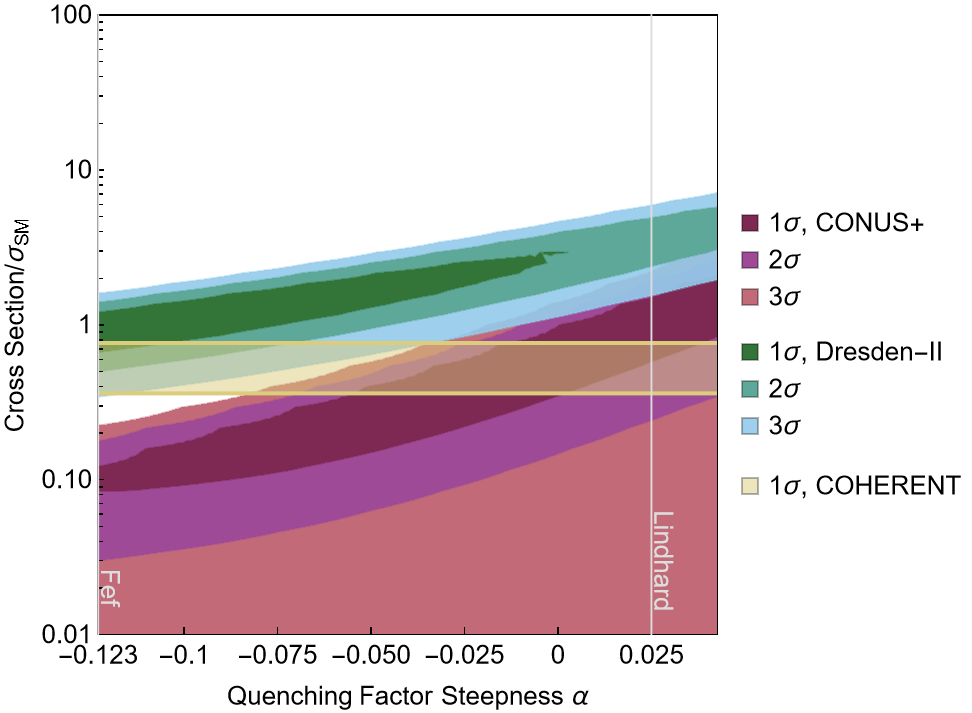}
     \includegraphics[width=0.495\linewidth]{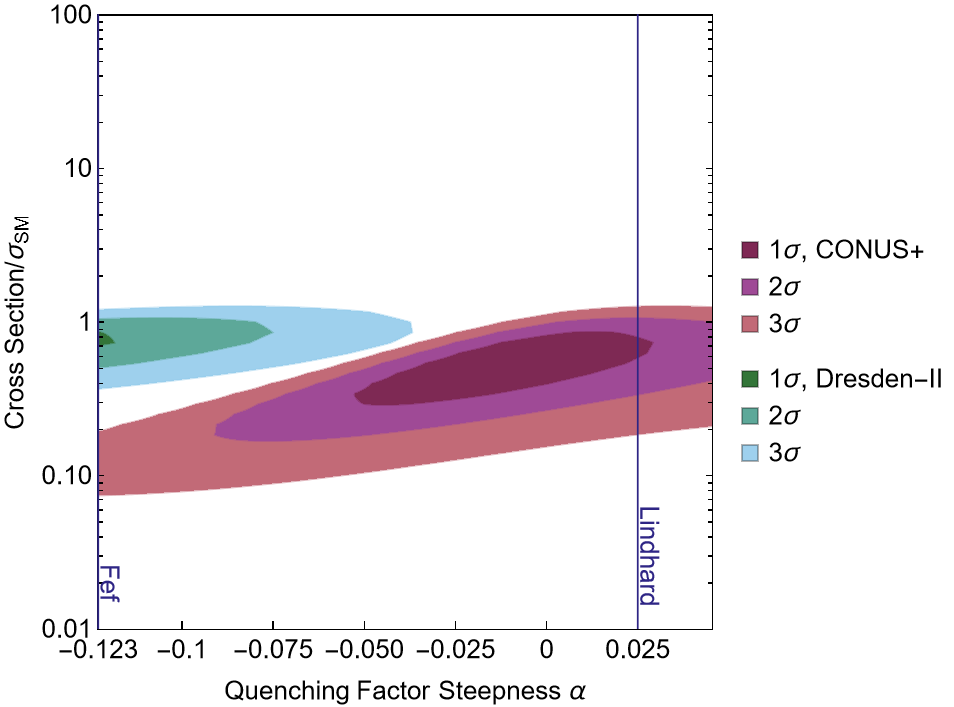}
     \includegraphics[width=0.495\linewidth]{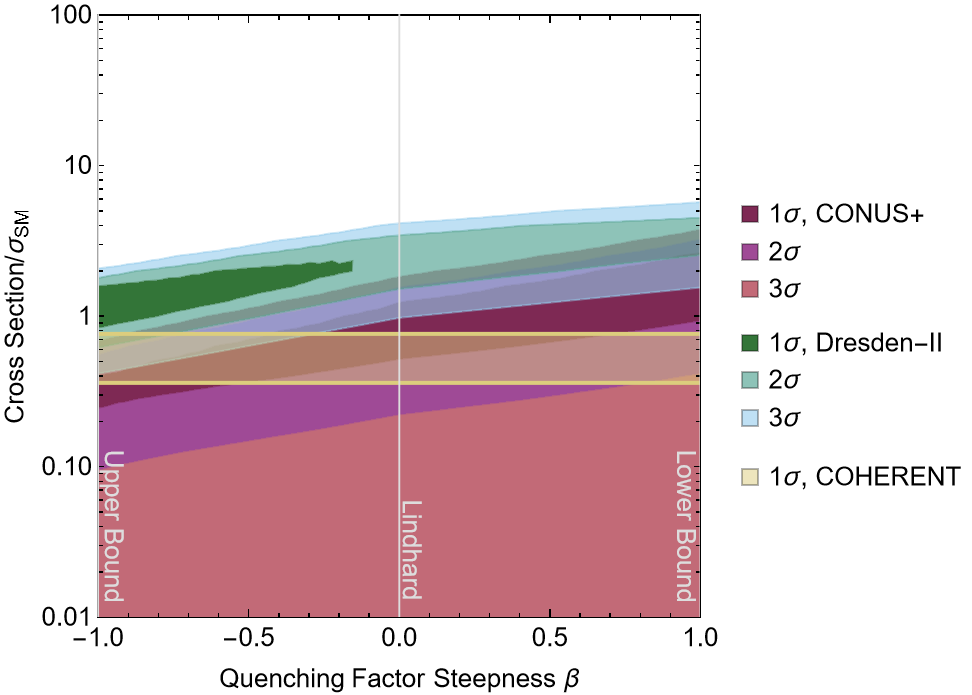}
     \includegraphics[width=0.495\linewidth]{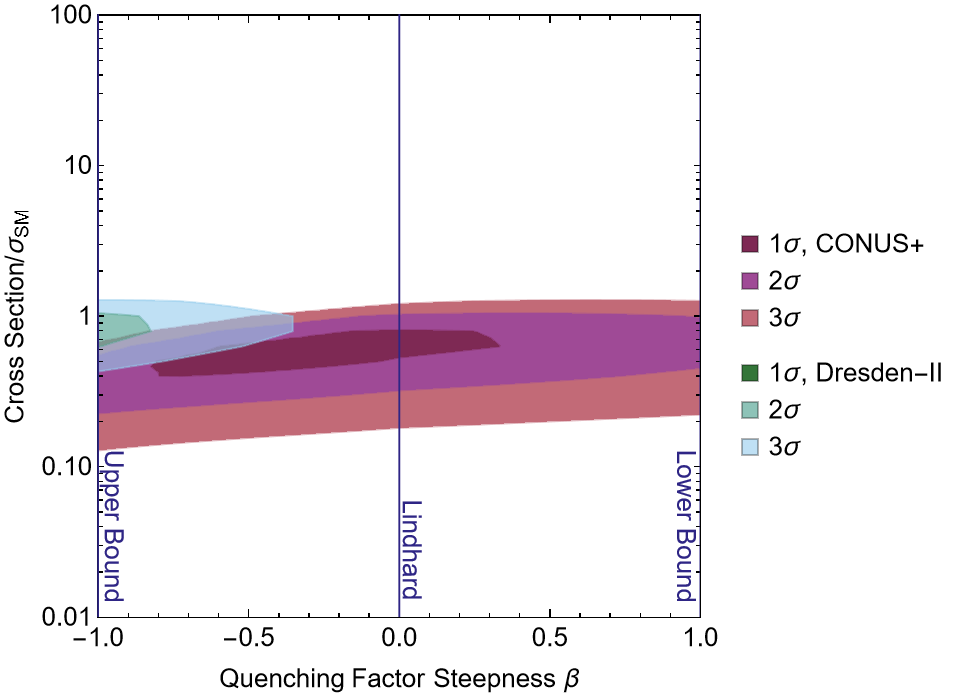}
    \caption{\textit{Upper left panel:} 1$\sigma$, 2$\sigma$ and 3$\sigma$ sensitivity contours to the ratio of CE$\nu$NS cross section normalization vs expectation in the SM, and quenching factor in germanium, for Dresden-II (green-blue palette) and CONUS+ (red-purple palette) and COHERENT (yellow). We further add vertical lines indicating the values of quenching factor steepness corresponding to the Lindhard model ($\alpha=0.025$) and FeF measurement ($\alpha=-0.123$), confer our parametrization from Eq. \ref{eq:qf_parametrization}. It can be clearly appreciated that CONUS+ and Dresden-II are largely discrepant for the quenching factor models considered. \textit{Upper right panel:} Combination of the COHERENT fit with Dresden-II and CONUS+ independently, at various sigmas. As apparent in the plot, the combination of Dresden-II and COHERENT is in significant tension with the combination of CONUS+ and COHERENT for all possible values of the quenching factor steepness. Interestingly, at present, the combination of CONUS+ and COHERENT favors a CE$\nu$NS cross section somewhat smaller than expected in the Standard Model, and a quenching factor in between the Fef measurement and the Lindhard model. \textit{Lower left panel:} Sensitivity contours to the ratio of of CE$\nu$NS cross section normalization vs expectation in the SM, and quenching factor in germanium, for Dresden-II (green-blue palette) and CONUS+ (red-purple palette), under the quenching factor parametrization from Eq. \ref{eq:qf_parametrization_2}. \textit{Lower right panel:} Combination of the COHERENT fit with CONUS+ and Dresden-II independently, for the quenching factor parametrization from \ref{eq:qf_parametrization_2}. Here we notice some overlap of the sensitivity contours within 3$\sigma$, unlike for the quenching factor parametrization from Eq. \ref{eq:qf_parametrization}.}
    \label{fig:CEnuNS_fit}
\end{figure*}
where the variance depends on the recoil energy $E_R$, and takes different values depending on the experiment under consideration. In Table \ref{tab:exp-params} we show the values used for the different experiments considered in this work, along with further relevant experimental details. At low recoil energies, the main source of uncertainty in the calculation of the recoil rates at a given experiment is given by the quenching factor. To tackle this source of uncertainty, we propose a parameterization that allows to interpolate among the most conservative (Lindhard model) and most aggressive (Fef measurements) scenarios. This is given by 
\begin{align}
&Q(E_R, \alpha, k) = 
    \begin{cases} 
        \alpha (E_R-1.73) +0.184, & E_R\leq 1.73 \, \rm keV \\
        Q_L(E_R, k) & E_R\geq 1.73 \,  \rm keV
    \end{cases}\\
    &Q_L(E_R, k)= \frac{k g(E_R)}{1+k g(E_R)}
    \label{eq:qf_parametrization}
\end{align}
where
\begin{equation}
\begin{aligned}
    g(E_R) &= 3(11.5 Z^{-7 / 3} E_R)^{0.15} + 0.7 (11.5 Z^{-7 / 3} E_R)^{0.6} \\
    &\quad + 11.5 Z^{-7 / 3} E_R
\end{aligned}
\end{equation}
where $E_R$ is the recoil energy in keV. $\alpha$ is the slope of the interpolating function at energies below $1.73 \pm 0.32 \rm keV$, and we take values in the range $\alpha = [-0.12,0.04]$. At energies above 1.73 keV, the Lindhard model and the Fef measurements are in good agreement. The Lindhard model presents another uncertain parameter, $k$, which in the following we fix to $k=0.162$. In addition, we propose yet another more conservative parametrization, which does not cover the Fef measurements, but rather is upper bounded by the upper limits derived by the CONUS+ collaboration together with the German National Metrology Institute (PTB) \cite{Bonhomme:2022lcz}. This parametrization still covers the lowest energy uncertain measurements from thermal neutron capture from \cite{Collar:2021fcl, Kavner:2024xxd, PhysRevC.4.125}. It is given by the following piece-wise function
\begin{widetext}
\begin{align*}
Q(E_R, k, \beta)=
\begin{cases}
Q_L(E_R, k) - \beta \left[ 0.1 + 0.25(E_R - 0.46) - Q_L(E_R, k) \right], & E_R < 0.6,\ \beta \leq 0 \\
Q_L(E_R, k) + \beta \left[ 0.28 - 0.3(E_R - 0.4) - Q_L(E_R, k) \right], & E_R < 0.6,\ \beta \geq 0 \\
Q_L(E_R, k)(1 + 0.17 \beta), & 0.6 < E_R < 1.73,\ \beta \leq 0 \\
Q_L(E_R, k)(1 + 0.30 \beta), & 0.6 < E_R < 1.73,\ \beta \geq 0 \\
Q_L(E_R, k), & E_R \geq 1.73
\end{cases}
\label{eq:qf_parametrization_2}
\end{align*}
\end{widetext}
where $\beta$ takes values in the range $\beta=(-1,1)$, and $Q_L(E_R, k)$ denotes the Lindhard quenching factor as described in Eq. \ref{eq:qf_parametrization}. We illustrate our proposed parameterizations of the quenching factor in Fig. \ref{fig:quenching_parametrization}, confronted with the extrapolated Lindhard model to low energies (solid pink line) and experimental measurements from iron-filtered monochromatic neutrons (Fef) and yttrium-berylium photoneutron source \cite{Collar:2021fcl,Scholz:2016qos}.
Following a nuclear recoil, an additional ionization rate is generated from the Migdal effect \cite{AtzoriCorona:2023ais,Herrera:2023xun, Maity:2024vkj,Liao:2021yog}, which can be obtained in detectable energy via
\begin{align}
    \frac{dR^{\rm mig}}{dE_{\rm det}} &= N_T \int dE_{R} \int dE_{er} \,
        \delta(E_{\rm det} - q_{\rm nr} E_{R} - E_{er} + |E^{nl}|) \nonumber \\
    &\times  \epsilon(E_{R}) \int_{E_\nu^{\min}}^{E_\nu^{\max}} \! dE_\nu \,
         \frac{\mathrm{d}\Phi_{\bar{\nu}_e}}{\mathrm{d} E_{\nu}} \frac{d\sigma}{dE_{R}} \times |Z_{\rm ion}(E_{er})|^2 \,.
\end{align}
The ionization form factor reads
\begin{align}
    |Z_{\rm ion}(E_{\rm er})|^2 = \frac{1}{2\pi} \sum_{n, l} p_{n l \to E_{\rm er}} \,,
\end{align}
and the differential transition probability of an electron in the orbital $(n, l)$ to an unbound state with recoil energy $E_{\rm er}$ is denoted by $p_{n l \to E_{\rm er}}$. We take the ionization probabilities for germanium from \cite{Ibe:2017yqa}.
The integration over nuclear recoils is performed in the kinematical range
\begin{equation}
\frac{\left(E_{\rm er}+|E^{nl}|\right)^2}{2 m_A}<E_{R}<\frac{\left(2 E_\nu-\left(E_{\rm er}+|E^{nl}|\right)\right)^2}{2\left(m_A+2 E_\nu\right)}.
\end{equation}
Finally, we include in our analysis contributions arising from elastic neutrino-electron scattering, although at the energies of interest such contributions are expected to be negligible compared to the quenched signal from CE$\nu$NS and the ionization signal from the Migdal effect. The differential scattering cross section for electron neutrinos on electrons reads
\begin{align}
    \frac{d\sigma}{dE_{er}} =
        \sum_{n,l} \theta(E_{\rm er} - |E^{nl}|)
        \frac{d\sigma^{0}}{dE_{\rm er}} ,
\end{align}
where the binding energies of electrons in the shells $(n,l)$ of germanium are included. The cross section for electron antineutrino scattering on free electrons ($\bar{\nu}_e + e \to \bar{\nu}_e+ e)$ given by \cite{Vogel:1989iv}
\begin{equation}
\begin{aligned}
\frac{\mathrm{d} \sigma^{0}}{\mathrm{~d} E_e} &= \frac{G_F^2 m_e}{2 \pi} \bigg[ (g_V+g_A)^2+(g_V-g_A)^2\left(1-\frac{E_{er}}{E_\nu}\right) \\
&\quad + (g_A^2-g_V^2) \frac{m_e E_{er}}{E_\nu^2} \bigg]
\end{aligned}
\end{equation}
where $g_V=\frac{1}{2}+2 \sin ^2 \theta_W$ and $g_A=-1/2$. The differential recoil rate is then
\begin{align}
    \frac{dR}{dE_{\rm er}} = \epsilon N_T \!\!
        \int_{E_\nu^{\min}}^{E_\nu^{\max}} \!\! dE_\nu \frac{\mathrm{d}\Phi_{\bar{\nu}_e}}{\mathrm{d} E_{\nu}}
        \frac{d\sigma}{dE_{\rm er}},
\label{eq:dRdE-sm}
\end{align}
with $\epsilon$ the efficiency of the experiment and $N_T$ denoting the number of target nuclei or electrons in the detector. The minimum neutrino energy required to induce a given recoil energy is now given by 
\begin{align}
    E_\nu^{\min} = \frac{E_{\rm er}
                 + \sqrt{2 m_e E_{\rm er} + E_{\rm er}^2}}{2} \,.
\end{align}
We will also assess the compatibility of these experimental results in a BSM scenario that is enhanced at low recoil energies, where uncertainties from the Ge quenching factor manifest. A standard example is a magnetic moment of the neutrinos, which is strongly suppressed in the minimally extended Standard Model with right handed neutrinos, but could be enhanced in BSM physics scenarios, see \textit{e.g} \cite{Babu:2020ivd, Lindner:2017uvt, Herrera:2025vjc}. The cross section from CE$\nu$NS reads
\begin{align}
\frac{\mathrm{d} \sigma_{\mu_\nu}}{\mathrm{~d} E_R}=Z^2\left(\frac{\mu_\nu}{\mu_B}\right)^2 \frac{\alpha^2 \pi}{m_e^2}\left[\frac{1}{E_R}-\frac{1}{E_\nu}\right]|F(E_R)|^2,
\end{align}
where the magnetic moment $\mu_{\nu}$ is given in units of the Bohr magneton \footnote{Reactor neutrino experiments are sensitive to an \textit{effective} neutrino magnetic moment comprising a sum of diagonal and off-diagonal contributions, defined as $\mu_{\nu}=\sum_k\left|\sum_j U_{\ell k}^* \mu_{j k}\right|^2$, where $U$ is the PMNS matrix.}. For elastic neutrino-electron scattering, the cross section reads
\begin{align}
\frac{\mathrm{d} \sigma_{\mu_\nu}}{\mathrm{~d} E_{\rm er}}=\left(\frac{\mu_\nu}{\mu_B}\right)^2 \frac{\alpha^2 \pi}{m_e^2}\left[\frac{1}{E_{\rm er}}-\frac{1}{E_\nu}\right].
\end{align}
\begin{figure*}[t!]
    \centering
    \includegraphics[width=0.49\linewidth]{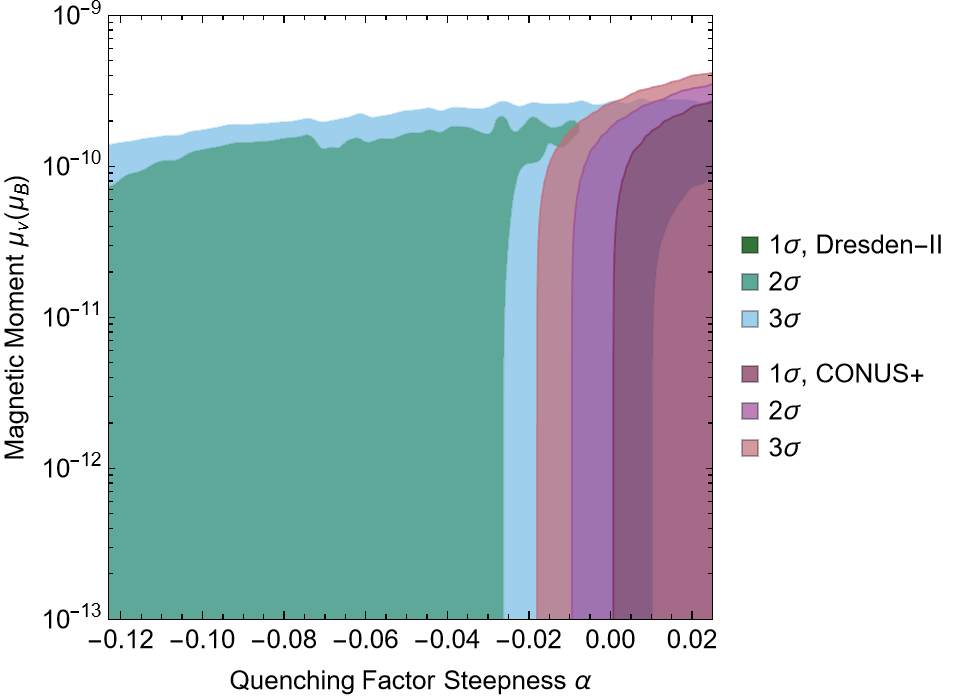}
    \includegraphics[width=0.49\linewidth]{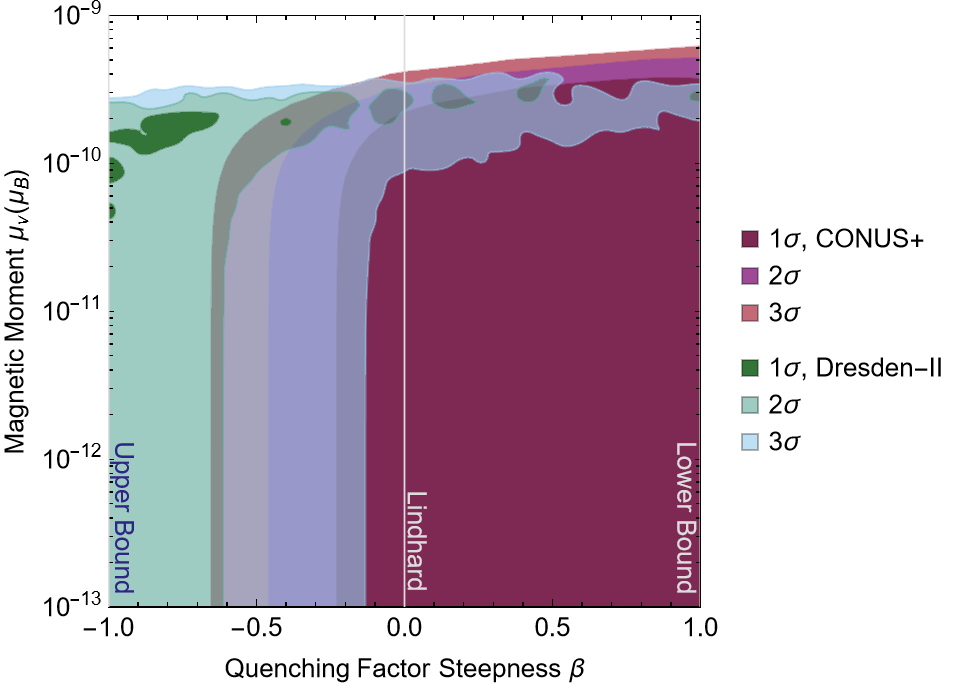}
    \caption{\textit{Left panel:} 1$\sigma$, 2$\sigma$ and 3$\sigma$ sensitivity contours on the neutrino magnetic moment and the Ge quenching factor (followinf the parametrization from Eq. \ref{eq:qf_parametrization}, for Dresden-II (green-blue palette) and CONUS+ (red-purple palette). In presence of a large magnetic moment, Dresden-II can accomodate the Lindhard model. On the other hand, when the magnetic moment is small, larger quenching factor as that from the Fef measurement is required to fit the data. CONUS+ strongly disfavors the Fef measurement, and is in agreement with the Lindhard model, albeit favoring a somewhat smaller value of the steepness parameter $\alpha$ (see main text for details). \textit{Right panel:} Analogous plot, but using the parametrization from Eq. \ref{eq:qf_parametrization_2} instead.}
    \label{fig:magnetic_moment_fit}
\end{figure*}

For illustration, we show in Figure \ref{fig:recoil_rate} the recoil rate in reconstructed (quenched) energy in germanium at Dresden-II, for CE$\nu$NS in the SM (solid black) and via a magnetic moment with strength $\mu_{\nu_{e}}=10^{-10} \mu_B$ (solid red), elastic neutrino-electron scattering in the SM (dashed black line) and via a magnetic moment (dashed red line) and via the Migdal effect (dotted green). The right panel corresponds to the quenching factor parametrization from Eq. \ref{eq:qf_parametrization}, and the right panel corresponds to Eq. \ref{eq:qf_parametrization_2}. For CE$\nu$NS in the SM, we also bracket the impact of quenching factor uncertainties with a colored band. It can be appreciated that the quenching factor uncertainties can affect the recoil rate at some reconstructed energies by up to a factor of $\sim 4$ for both parametrizations chosen, while the effect of quenching factor uncertainties at large recoil differs, being more relevant for the parametrization from Eq.~\ref{eq:qf_parametrization}. The contribution from the Migdal effect is negligible below $0.4 \,  \text{keV}_{ee}$, but is sizable and even dominant at the largest recoil energies measured by reactor neutrino experiments. As expected, the magnetic moment enhances the recoil rate towards low energies, with somewhat steeper index than the recoil rate arising from CE$\nu$NS in the SM.

\section{Combined analysis of CE$\nu$NS cross section and quenching factor}
Following the description of the recoil rate calculation from previous section, we can derive sensitivity contours from Dresden-II and CONUS+ in the parameter space spanned by the ratio of phenomenological cross section vs SM expected cross section (confer Eq. \ref{eq:xsec-sm-7s}) and the quenching factor steepness parameter $\alpha$, confer Eq. \ref{eq:qf_parametrization}. For this purpose, we perform a binned Poisson maximum likelihood estimation. First, we define the recoil rate with two parameters: $dR(n,\alpha)/dE_\mathrm{det}$, where $n=\sigma'/\sigma_{\mathrm{SM}}$, is the magnitude of the cross-section with respect to SM cross-section $\sigma_\mathrm{SM}$, and the quenching factor parameter $\alpha$. We then define the $\chi^2$ function for a Poissonian distribution
\begin{align}
\chi^2=\sum_{i=1}^b \bigg(2[ \langle x_i \rangle -x_i+2x_i\log\frac{x_i}{\langle x_i \rangle}\bigg),
\end{align}
where $b$ denotes the number of bins. Both CONUS+ and Dresden-II have $10$ eV sized bins. $\langle x_i \rangle=dR_i(n,\alpha)/dE_\mathrm{det}+bg_i$ corresponds to the expectation value of events in each bin $i$, and $x_i=s_i+bg_i$ corresponds to the sum of signal events $s_i$ and background events $bg_i$. For both Dresden-II and CONUS+, we used reactor-ON data for $x_i$ and reactor-OFF data for $bg_i$. We include all contributions arising from CE$\nu$NS, the Migdal effect and neutrino-electron scattering. The fits are largely dominated by CE$\nu$NS, but note that the contribution from the Migdal effect is not negligible and amounts to about $\sim 5 \%$ in total.

In Figure \ref{fig:CEnuNS_fit}, we show sensitivity contours on the CE$\nu$NS cross-section and germanium quenching factor at low energies from CONUS+ (red-purple palette) and Dresden-II (green-blue palette). In the upper panels of the Figure, the quenching steepness parameter at $\alpha=-0.123$ corresponds to a linear regression fitting of the Iron-filter (Fef) measurements, and at $\alpha=0.025$ corresponds to a linear fitting of the Lindhard model, confer Eq. \ref{eq:qf_parametrization}. For the lower panels, we follow the parametrization inspired by the CONUS+ measurements \cite{Bonhomme:2022lcz}, confer Eq. \ref{eq:qf_parametrization_2}.
At the SM CE$\nu$NS cross-section (corresponding to a ratio equal to 1 in the y-axis of the Figure), Dresden-II favors the Fef quenching factor measurement, whereas CONUS+ favors a value of the quenching factor somewhat smaller or equal than the Lindhard model. We find that Dresden-II is incompatible with the Lindhard model at $3.79\sigma$, and CONUS+ is in tension with the Fef measurement at huge significance (we report $\Delta\chi^2=410.9$). We also compute the tension among CONUS+ and the best fit quenching factor of Dresden-II, finding 18.5 $\sigma$. For the tension among Dresden-II and the best-fit quenching factor from CONUS+, we find 4.37 $\sigma$. To calculate the tension between the two experiments using the so-called parameter goodness of fit, we follow \cite{Maltoni:2003cu, Chatterjee:2024kbn}, and define $\Delta\chi^2_{\mathrm{tot}}=\chi^2_{\mathrm{Dresden-II}}+\chi^2_{\rm CONUS+}-\sum \chi^2_{\mathrm{min}}$. Then, the $\Delta\chi^2_{\mathrm{tot, min}}$ would be the test statistics for goodness-of-fit evaluation, which in our case is $\Delta\chi^2_{\mathrm{tot, min}}=14.39$. With this prescription, we find a $3.47\sigma$ tension between Dresden-II and CONUS+.
As can be seen in the left-hand panels of Fig. \ref{fig:CEnuNS_fit}, we placed a $1\sigma$ contour band from the COHERENT measurement of CE$\nu$NS in germanium \cite{Adamski:2024yqt}. The COHERENT band is flat, since this experiment is a stopped-pion neutrino source which involves higher neutrino energies than reactor neutrino experiments, and the quenching factor is in agreement with the Lindhard model. The form factor we choose for COHERENT analysis is the Klein-Nystrand model \cite{Klein:1999qj}.  We further add the $\Delta\chi^2$ of COHERENT to the Dresden-II and CONUS+ sensitivity in the right hand panels of Figure \ref{fig:CEnuNS_fit}. For the parametrization in \ref{eq:qf_parametrization}, we find that the inclusion of COHERENT increases the tension between CONUS+ and Dresden-II, beyond 3$\sigma$ regardless of the quenching factor considered, and favors a somewhat smaller cross section than expected in the SM at CONUS+. For the parametrization in \ref{eq:qf_parametrization_2}, however, we find a milder tension between CONUS+ and Dresden, closer to the $2\sigma$ level.

\section{Combined analysis of neutrino magnetic moment, CE$\nu$NS cross section and quenching factor}
We further derive sensitivity contours on the neutrino magnetic moment vs quenching factor steepness at CONUS+ and Dresden-II, see Figure \ref{fig:magnetic_moment_fit}. We fix the CE$\nu$NS cross section to the expected value in the SM, and focus on the incoherent contribution induced by the neutrino magnetic moment at low recoil energies. In this analysis we ignore COHERENT data, since we have fixed the cross section to the SM value and quenching factor uncertainties are negligible at COHERENT. As in the analysis of the CE$\nu$NS cross section, we include all contributions arising from CE$\nu$NS, the Migdal effect and neutrino-electron scattering. Again, fits are dominated by CE$\nu$NS, although the contributions from the Migdal effect and electron scattering are significantly larger in presence of a neutrino magnetic moment, of about $\sim 33 \%$. This is in concordance with the recoil rate calculations shown in Fig. \ref{fig:recoil_rate}.
\begin{figure*}[t!]
    \centering
    \includegraphics[width=0.49\linewidth]{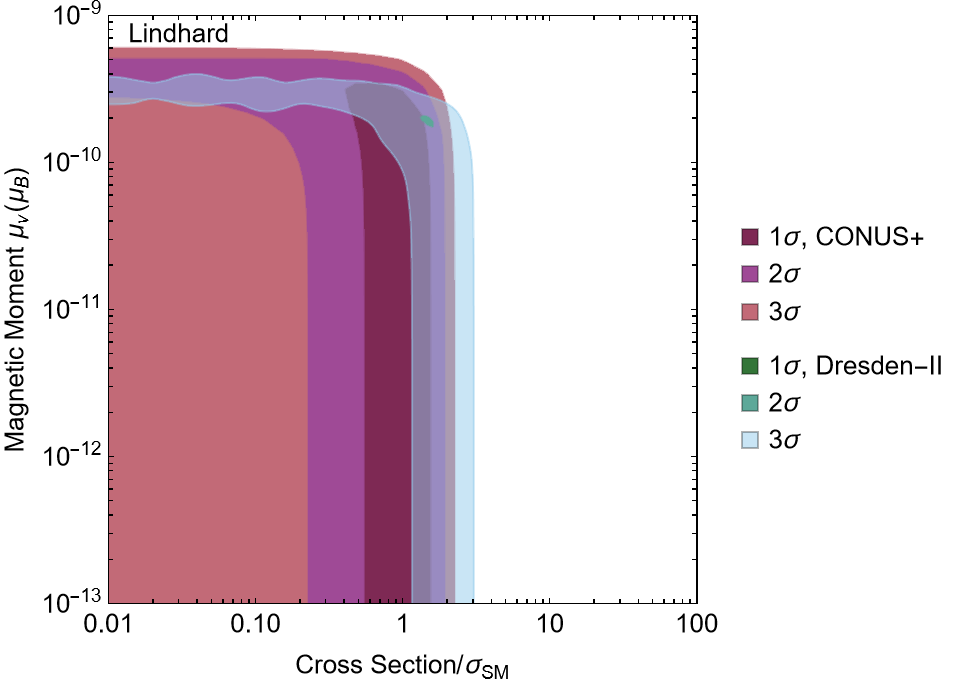}
    \includegraphics[width=0.49\linewidth]{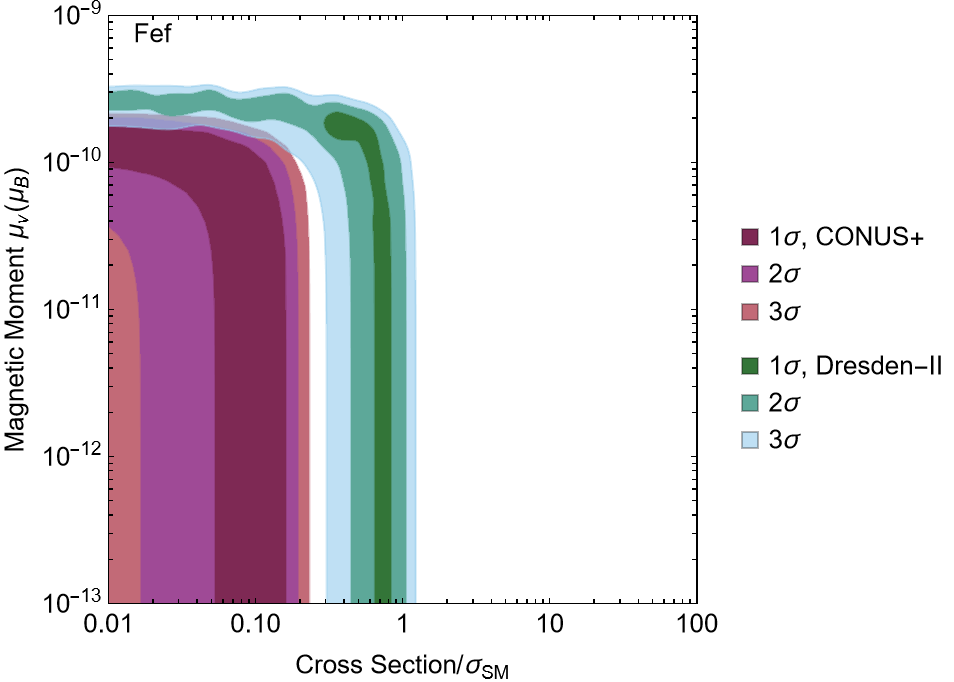}
     
    \caption{\textit{Left panel:} 3$\sigma$ contour from Dresden-II on the parameter space spanned by the CE$\nu$NS cross section ratio to the SM expectation and the neutrino magnetic moment, for the quenching factor from the Lindhard model. If fixing the cross section to the SM expectation and the quenching factor to the Lindhard model, Dresden-II would indicate a preference for a large magnetic moment. If the magnetic moment is small, however, the Lindhard model enforces the cross section to be somewhat larger than the SM expectation. \textit{Right panel}: Same as the left panel, but fixing the quenching factor to the Fef measurement instead. In this case, there is no combination of magnetic moment and CE$\nu$NS cross section normalization that allows to reconcile both data sets well.}
    \label{fig:CEnuNS_vs_magnetic}
\end{figure*}
In Figure \ref{fig:magnetic_moment_fit}, we show sensitivity contours on the electron neutrino magnetic moment and germanium quenching factor at low energies, when assuming that the SM cross section for CE$\nu$NS aligns with the theoretical prediction from Eq. \ref{eq:xsec-sm-7s}. The left panel considers the parametrization from \ref{eq:qf_parametrization}, while the right panel considers the parametrization from \ref{eq:qf_parametrization_2}. As expected, CONUS+ excludes small values of the quenching factor parameters $\alpha$ and $\beta$, corresponding to large values of the quenching factor at low energies. Regardless of the value of the magnetic moment of neutrinos, if such quenching factor were present, the total number of recoil events expected from CE$\nu$NS in the SM would surpass the observed number of events by the collaboration. On the other hand, Dresden-II can't exclude small values of the quenching factor parameters $\alpha$ and $\beta$ (large values of the quenching factor), since those align with the observed number of the events by the collaboration when considering the expected CE$\nu$NS cross section in the SM. Indeed, for small values of $\alpha$ and $\beta$, the constraint on the neutrino magnetic moment becomes more stringent than at high values of. Interestingly, we find that a quenching factor in between that as extracted from the Fef measurements and the Lindhard model, and a neutrino magnetic moment in the range of $10^{-10} \mu_{B}$ is preferred by Dresden-II data. That region of parameter space is, however, in tension with several other astrophysical \cite{Raffelt:1990pj,Capozzi:2020cbu, Arceo-Diaz:2015pva}, cosmological \cite{Morgan:1981zy,Elmfors:1997tt,Carenza:2022ngg, Li:2022dkc}, and laboratory constraints \cite{LZ:2023poo,XENON:2022ltv, Beda:2013mta, Akhmedov:2022txm, Borexino:2017fbd}. Finally, we note that the sensitivity of CONUS+ to the neutrino magnetic moment can vary by up to one order of magnitude in the quenching factor steepness range going from $\alpha \sim (-0.02,0.03)$, or $\beta \sim (-0.6,0)$. It is worth mentioning that under the parametrization from Eq. \ref{eq:qf_parametrization_2}, the overlap of the CONUS+ and Dresden-II is more significant, but there is still a discrepancy a the $1\sigma$ level in the preferred parameter space of the magnetic moment.

Finally, in Figure \ref{fig:CEnuNS_vs_magnetic} we show sensitivity contours from CONUS+ and Dresden-II in the parameter space spanned by the effective reactor neutrino magnetic moment and the CE$\nu$NS cross section, for fixed choices of the quenching factor: Lindhard (left panel) and Fef (right panel). Interestingly, we find that there are combinations of the magnetic moment and cross section normalization that allow to reconcile CONUS+ and Dresden-II for the Lindhard model, while there is only a poor overlap for the Fef measurement. Furthermore, the relative strength of CONUS+ and Dresden-II bounds on the neutrino magnetic moment is significantly affected by the choice of quenching factor. In particular, for the Lindhard model, the sensitivity of Dresden-II to the neutrino magnetic moment is a factor of $\sim 2$ stronger than that from CONUS+, while for the Fef measurement the sensitivity of CONUS+ is stronger than Dresden-II by a factor of $\sim 3$. We highlight that current uncertainties on the quenching factor prevent us from claiming any deviations induced by a magnetic moment or a different normalization of the CE$\nu$NS cross section than expected in the SM. The sensitivity contours can shift to either side of the SM expected cross section depending on the choice of quenching factor, and the degeneracy of the CE$\nu$NS normalization with $\textit{e.g}$ a neutrino magnetic moment is quite significant.

\section{Conclusions}
It is of outmost importance to reduce uncertainties on quenching factors at sub-keV recoil energies if aiming to detect new physics from neutrinos at the low-energy frontier. Several measurements of/limits on  CE$\nu$NS from different experiments have been reported in recent years, some of which rely on different choices of the quenching factor in germanium and favor (albeit at small significance) deviations from the expected SM cross section. 

We have quantified these previous intuitions in the literature and reported progress in this regard, by combining the data from Dresden-II, CONUS+ and COHERENT accounting for uncertainties on the quenching factor of germanium. We have proposed two novel analytical parametrizations of the quenching factor (confer Eq. \ref{eq:qf_parametrization} and Eq. \ref{eq:qf_parametrization_2}) that interpolate among the Fef measurement and the Lindhard model, and among the lower and upper errors of the CONUS+ and PTB measurement \cite{Bonhomme:2022lcz}. By doing this, we were able to derive sensitivity contours of these experiments on the ratio of the CE$\nu$NS cross section to the SM expectation and the quenching factor steepness at energies below 1.73 keV, where the Fef measurement clearly differs from the Lindhard model.

We found that the combination of Dresden-II and COHERENT favors the expected CE$\nu$NS cross section in the SM and a quenching factor closer to the Fef measurement. We note though that values of the quenching factor steepness in the ranges $\alpha \sim (-0.12,-0.03)$ or $\beta \sim (-1,-0.4)$ are allowed within $3 \sigma$, see Fig. \ref{fig:CEnuNS_fit}. On the other hand, the combination of COHERENT and CONUS+ prefers a cross section a factor of $\sim 2$ smaller than that expected in the SM, while indicating a quenching factor somewhat smaller than the Lindhard model ($\alpha \sim (-0.05,0.25)$ or $\beta \sim (-0.8,0.3)$ at 1$\sigma$). Interestingly, a the $3 \sigma$ level CONUS+ could still be reconcilable with the Fef measurement, but the cross section should be suppressed w.r.t to the SM by a factor of $\sim 10$.

We have further derived the sensitivity of these experiments to the effective reactor neutrino magnetic moment and quenching factor steepness. We find that Dresden-II data may be compatible with a quenching factor steepness closer to the Lindhard model, in presence of a large neutrino magnetic moment at the level of $\mu_{\nu} \sim 10^{-10} \mu_{B}$. Moreover, the sensitivity of Dresden-II and CONUS+ is significantly affected by the choice of quenching factor, changing $\textit{e.g}$ the constraints in CONUS+ by up to one order of magnitude. When fitting the combination of CE$\nu$NS cross section normalization and magnetic moment for fixed choices of the quenching factor, we find that the Lindhard model can better reconcile CONUS+ and Dresden-II than than the Fef measurement.

Future experiments with larger exposure, lower energy threshold and lower backgrounds in germanium may shed further light on these apparent discrepancies. A more dedicated effort to measure the quenching factors independently from neutron scattering may allow to reduce uncertainties on the recoil rate, which will help to a better measurement of the CE$\nu$NS cross section at low energies and to be able to claim fainter signals induced by new physics. Measurements on complementary targets to germanium such as CaWO$_4$ will also be useful in this regard, see \cite{Thulliez:2020esw,CRAB:2022rcm,CRESST:2023cxk,Kluck:2024sao}. We hope that our work will allow for a better assessment of the quenching factor uncertainties in future CE$\nu$NS studies, which we have demonstrated that play a crucial role in the interpretation of current experimental results.

\begin{section}{ACKNOWLEDGEMENTS} 

We thank J. Hakenm\"uller for information about the CONUS and CONUS+ experiments and data. We also thank M. Lindner for drawing our attention to Ref.~\cite{Bonhomme:2022lcz}.
This work was supported by award DE-SC0020262 of the U.S. Department of Energy Office of Science.

\end{section}

\bibliography{References}    

\clearpage

\end{document}